\documentclass[letterpaper,10pt]{article} 

\usepackage{opticameet3} 
\usepackage[nohyperlinks,nolist]{acronym}
\acrodef{AWGN}[AWGN]{additive white Gaussian noise}

\acrodef{BMI}[BMI]{binary mutual information}
\acrodef{BPS}[BPS]{blind phase search}

\acrodef{CD}[CD]{chromatic dispersion}
\acrodef{CMA}[CMA]{constant-modulus algorithm}
\acrodef{CNN}[CNN]{convolutional neural network}
\acrodef{CPE}[CPE]{carrier-phase estimation}

\acrodef{DFE}[DFE]{decision feedback equalizer}
\acrodef{DP}[DP]{dual-polarization}
\acrodef{DSP}[DSP]{digital signal processing}

\acrodef{ELU}[ELU]{exponential linear unit}
\acrodef{ELBO}[ELBO]{evidence lower bound}

\acrodef{FEC}[FEC]{forward error correction}
\acrodef{FIR}[FIR]{finite impulse response}

\acrodef{GD}[GD]{gradient descent}

\acrodef{IP}[IR]{impulse response}
\acrodef{ISI}[ISI]{inter-symbol interference}

\acrodef{KL}[KL]{Kullback-Leibler}
\acrodef{LDPC}[LDPC]{low-density parity-check}

\acrodef{MA}[MA]{moving average}
\acrodef{MAP}[MAP]{maximum a posteriori}
\acrodef{MIMO}[MIMO]{multiple-input multiple-output}
\acrodef{ML}[ML]{maximum likelihood}
\acrodef{MMA}[MMA]{multi-modulus algorithm}
\acrodef{MMSE}[MMSE]{minimum mean squared error}

\acrodef{NN}[NN]{neural network}

\acrodef{PCS}[PCS]{probabilistic constellation shaping}
\acrodef{pdf}[pdf]{probability density function}
\acrodef{PMD}[PMD]{polarization mode dispersion}
\acrodef{pmf}[pmf]{probability mass function}
\acrodef{PSK}[PSK]{phase shift keying}
\acrodef{PSP}[PSP]{principal state of polarization}

\acrodef{QAM}[QAM]{quadrature amplitude modulation}

\acrodef{RDE}[RDE]{radius-directed equalizer}
\acrodef{RRC}[RRC]{root-raised cosine}

\acrodef{SER}[SER]{symbol error rate}
\acrodef{SNR}[SNR]{signal-to-noise ratio}
\acrodef{sps}[sps]{samples per symbol}

\acrodef{VAE}[VAE]{variational autoencoder}
\usepackage{wrapfig}

\usepackage{bm}

\usepackage{mleftright}

\usepackage[colorlinks=true,bookmarks=false,citecolor=blue,urlcolor=blue]{hyperref} 

\usepackage{siunitx}
\newcommand{\Meq}{M_{\mathrm{eq}}}
\newcommand{\Mest}{M_{\mathrm{est}}}

\newcommand{\elr}{\epsilon_{\mathrm{lr}}}
\newcommand{\Nb}{N_{\mathrm{batch}}}	
\newcommand{\taupmd}{\tau_{\mathrm{pmd}}}
\newcommand{\Lcd}{L_{\mathrm{cd}}}

\newcommand{\gammahv}{\gamma_{\mathrm{hv}}}

\newcommand{\lr}[1]{\left(#1\right)}
\newcommand{\mlr}[1]{\mleft(#1\mright)}

\def\e{{\mathrm{e}}}

\newcommand\authormark[1]{\textsuperscript{#1}}

\begin{document}

\title{Improving the Bootstrap of Blind Equalizers with Variational Autoencoders}

\author{Vincent Lauinger,\authormark{1,*} Fred Buchali,\authormark{2} and Laurent Schmalen\authormark{1}}

\address{\authormark{1}Communications Engineering Lab (CEL), Karlsruhe Institute of Technology (KIT), 76131 Karlsruhe, Germany\\
\authormark{2}Nokia, 70469 Stuttgart, Germany } 

\email{\authormark{*}\texttt{vincent.lauinger@kit.edu}}

\begin{abstract}
We evaluate the start-up of blind equalizers at critical working points, analyze the advantages and obstacles of commonly-used algorithms, and demonstrate how the recently-proposed \ac*{VAE} based equalizers can improve bootstrapping.
\end{abstract}
\vspace*{-0.0ex}
\section{Introduction}\vspace*{-1.5ex}
	In coherent optical communications, there is a great demand for flexible receiver algorithms which operate blindly and adapt to varying channel conditions. Operating receivers without known pilot symbols increases the effective data rate, which can be used to increase the forward error correction (FEC) overhead (and hence the robustness) or the net data throughput. However, the design of blind receivers is challenging, especially with modern higher-order modulation formats and \ac{PCS}. 
	In particular, the bootstrapping of blind algorithms, especially adaptive equalizers, is a major hurdle which has not been widely studied in the literature and on which we focus in this work. 
	
	A widely-used algorithm for bootstrapping blind equalization is the \ac{CMA}~\cite{godard}. Although designed for modulation formats with constant signal amplitude, it also converges for multi-amplitude formats such as $M$-QAM, where its criterion is ill-matched. 
	There are variants of the \ac{CMA} for multi-amplitude signals, e.g., the \ac{RDE}~\cite{ready_RDE}, which, however, have worse convergence characteristics at startup, when the correct radii cannot be estimated reliably. Hence, the standard \ac{CMA} is often used for bootstrapping, and the algorithms are switched or the amount of radii is increased once the \ac{CMA} has converged~\cite{guiomar_fully_blind_2015,Savory_AS_2010digital}. 
	However, the \ac{CMA} struggles with \ac{PCS} formats~\cite{zervas1991effects} or at certain working points. In this work, we compare different implementation variants of the \ac{CMA} at those critical points with the recently-proposed \ac{VAE} based equalizer~\cite{lauinger2022blind}. 
	The latter explicitly considers the \textit{a~priori} density of the transmitted symbols, which can prevent the equalizer to converge to invalid constellations. 
 
    \vspace*{-1ex}

\section{Blind Equalization} \vspace*{-1.5ex}
\begin{wrapfigure}[10]{R}{0.37\textwidth}
	\vspace*{-7ex}
	\raggedright
	\includegraphics{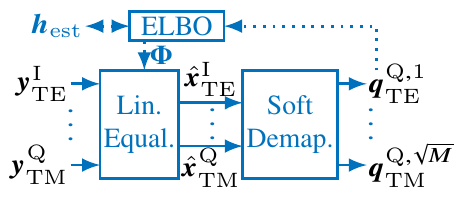} \\%
	\includegraphics{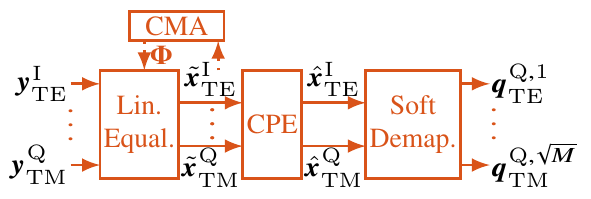} \\%
	\vspace*{-2.5ex}
	\caption{Sketch of VAE (top) and CMA (bottom).} \label{Eq_structure}
\end{wrapfigure}
The widely-used equalizer for bootstrapping a coherent optical transmission system is the \ac{CMA}~\cite{Savory_AS_2010digital}. In its common implementation~\cite{godard}, the distance of the equalized signal amplitude to the kurtosis of the transmitted constellation is minimized by updating the equalizer taps \emph{symbol-wise}. By nature, the \ac{CMA} is phase-insensitive and requires an additional \ac{CPE}, where we apply the blind \emph{Viterbi-Viterbi} algorithm~\cite{viterbi1983nonlinear} with a sufficient averaging over 501~symbols. Furthermore, there are variants of the CMA applying batch-wise updating via \ac{GD}, e.g., as proposed in~\cite{crivelli2014}, which we denote as \emph{CMAbatch}. 

In comparison, we analyze the recently proposed \ac{VAE} based equalizers~\cite{lauinger2022blind}, which use the \ac{ELBO} to approximate \ac{ML} channel estimation and, as a byproduct, equalize the received signal. 
The structures of both equalizer types are shown in Fig.~\ref{Eq_structure}. 
All algorithms update a similar linear equalizer block consisting of a complex-valued $2\times2$ butterfly structure with \ac{FIR} filters of length $\Meq$, 
and we use a maximum \textit{a~posteriori} based soft demapper as in \cite{cho2019probabilistic}.
In fact, the \ac{VAE} does not require a \ac{CPE}\footnote{Note that in the presence of laser phase noise, an additional \ac{CPE} might be necessary if the VAE cannot adapt rapidly enough.} 
and jointly updates---along with the equalizer taps---a similar but separate butterfly structure $\bm{h}_\text{est}$ of $\Mest$-long \ac{FIR} filters for channel estimation. We use an Adam optimizer to update either batch-wise (\emph{VAEbatch}) or with the \emph{VAEflex} scheme as proposed in~\cite{lauinger2022blind}. 
The latter is based on batch-wise processing, but instead of equalizing the whole batch per iteration, it only equalizes $N_\text{flex} <\Nb$~symbols. In the next iteration, it equalizes the next $N_\text{flex}$~symbols, so it re-uses $\Nb -N_\text{flex}$~symbols per iteration which boosts convergence. Further details can be found in~\cite{lauinger2022blind}. We also propose the \emph{CMAflex} scheme, which applies the same scheme based on the CMAbatch. 

The batch-wise schemes consider a batch of $\Nb = 200$~symbols with learning rates of $\elr=\num{1.2e-4}$ (CMAbatch) and $\elr=\num{2e-3}$ (VAEbatch); the CMA uses $\elr=\num{8e-4}$. The CMAflex/VAEflex schemes equalize $N_\text{flex} = 10$~symbols while processing $\Nb = 100$~symbols per iteration, and use {$\elr=\num{4.5e-5}$}~(CMAflex) and $\elr=\num{2e-3}$ (VAEflex). All hyperparameters are optimized individually for each scheme to reduce the percentage of failed runs during bootstrapping. 

\begin{figure} [!b]
		\centering
		\vspace*{-2ex}
		\includegraphics{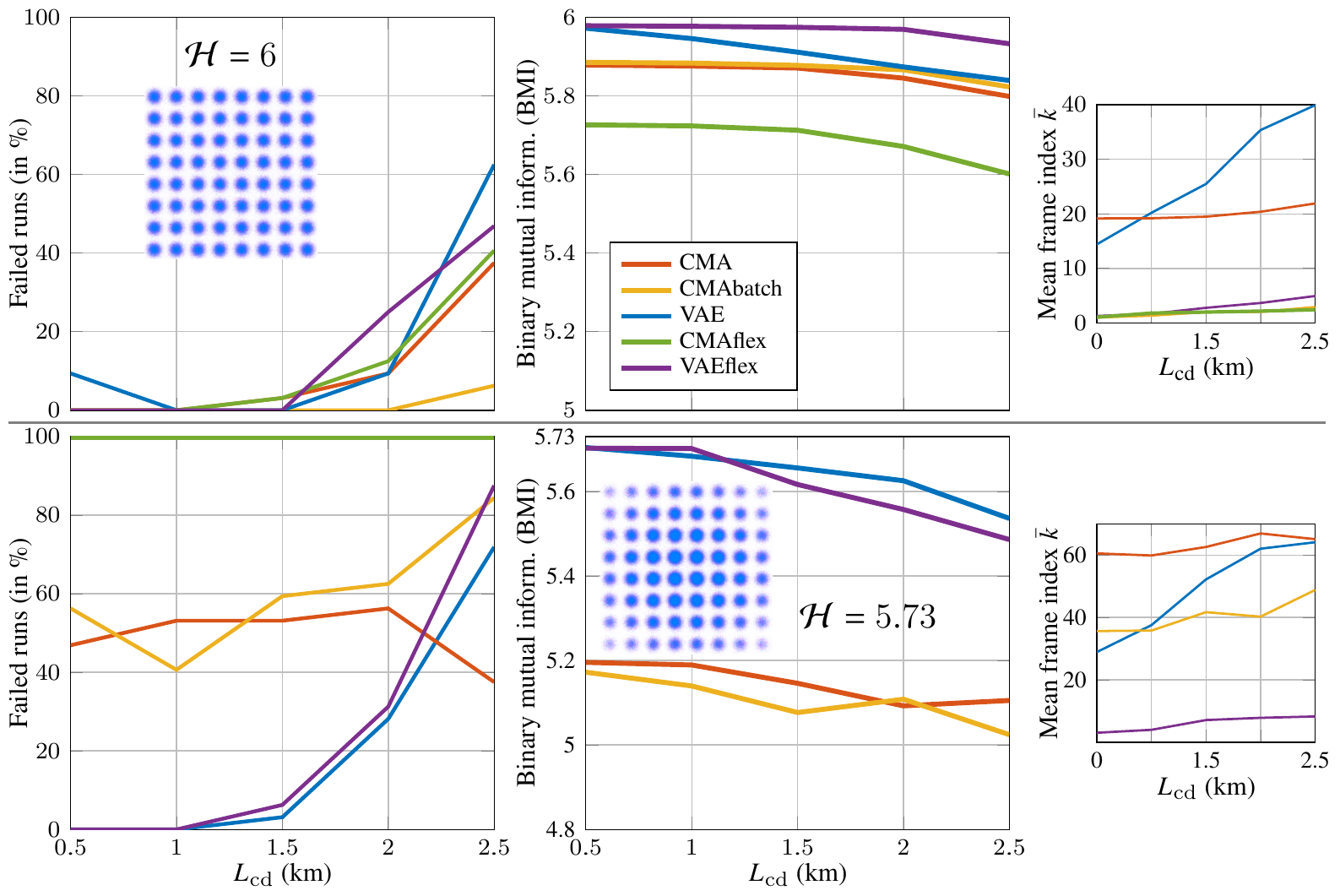}
		\vspace*{-4.5ex}
		\caption{Results for different residual \ac*{CD}: the top plots correspond to unshaped $64$-QAM ($\mathcal{H}=6$), the bottom plots to \ac*{PCS}-$64$-QAM ($\mathcal{H}=5.73$). The inset constellation diagrams show the equalized signal after the VAEbatch at $\Lcd=\SI{1.5}{\kilo\meter}$. The plots on the right show the mean frame index $\bar{k}$ at which a threshold of $\text{BMI}_\text{thr}=5$ and $\text{BMI}_\text{thr}=4.8$ (for the \ac*{PCS}, respectively) is reached.  }

		\label{Results:Lcd_full}
\end{figure} 
\vspace*{-1ex}
\section{Simulation Environment}\vspace*{-1.5ex}
\def\j{{\mathrm{j}}}
To analyze the bootstrapping of the different equalizers, we adapt the simulation environment of \cite{lauinger2022blind} (see also references therein) and use the same averaging scheme. 
Precisely, we transmit uniform $64$-QAM and \ac{PCS}-$64$-QAM with an entropy of $\mathcal{H}=5.73$ bits per symbol (and a Maxwell-Boltzmann distribution), assume a time-invariant channel during the bootstrapping phase, and model the fiber by the linear frequency domain channel matrix \vspace*{-0.7ex}
\begin{align*}
	\bm{H}\lr{f} &= \bm{R} \begin{pmatrix} \e^{\j\pi \tau_{\text{pmd}}  f } & 0 \\ 0 & \e^{-\j\pi \tau_{\text{pmd}}  f } \end{pmatrix}  \e^{-\j2\pi^2 \beta_{\text{cd}} L_{\text{cd}}  f^2 }  , \qquad \text{with} \quad \bm{R} = \begin{pmatrix} \cos\mlr{\gamma_{\text{hv}}} & \sin\mlr{\gamma_{\text{hv}}} \\ -\sin\mlr{\gamma_{\text{hv}}} & \cos\mlr{\gamma_{\text{hv}}} \end{pmatrix} \ ,
\end{align*}
where $\bm{R}$ represents a static rotation of the reference polarization to the fiber's \ac{PSP}, called HV-phase-shift $\gamma_{\text{hv}}$. In contrast to \cite{lauinger2022blind}, we do not consider the initial HV-phase-shift as compensated yet. 
Additionally, we include first-order \ac{PMD} caused by the differential group delay $\taupmd$ between the \acp{PSP},
and residual \ac{CD}, which is defined by the fiber's group velocity dispersion (GVD) parameter $\beta_{\text{cd}}=-\SI{26}{\pico\second^2\per\kilo\meter}$ (equals $D_{\text{cd}} = \SI{20}{\pico\s\per\nano\m\per\kilo\m}$ at $\lambda=\SI{1550}{\nano\m}$) times the uncompensated fiber length $\Lcd$. 
Complex \Ac{AWGN} is added on both polarizations. 

We evaluate $N_\text{ind}=100$~frames ($N_\text{ind}=20$ for the VAEflex/CMAflex schemes) 
and apply the same averaging scheme as in \cite{lauinger2022blind}, which results in an averaging over $80\,000\ldots100\,000$ symbols per frame index and run. 
As performance metric, we use both the \ac{BMI} (also often denotes as GMI) and  the percentage of \emph{failed} runs, which we define as runs where $\text{BMI}<5$ ($\text{BMI}<4.8$ for \ac{PCS} with $\mathcal{H}=5.73$, respectively) after a sufficient number of samples.
Similarly to~\cite{lauinger2022blind}, we apply a scheduler which halves the learning rate after every 20 (5 for the VAEflex/CMAflex schemes) frames for all algorithms.

\begin{figure} [!t]
		\centering
		\includegraphics{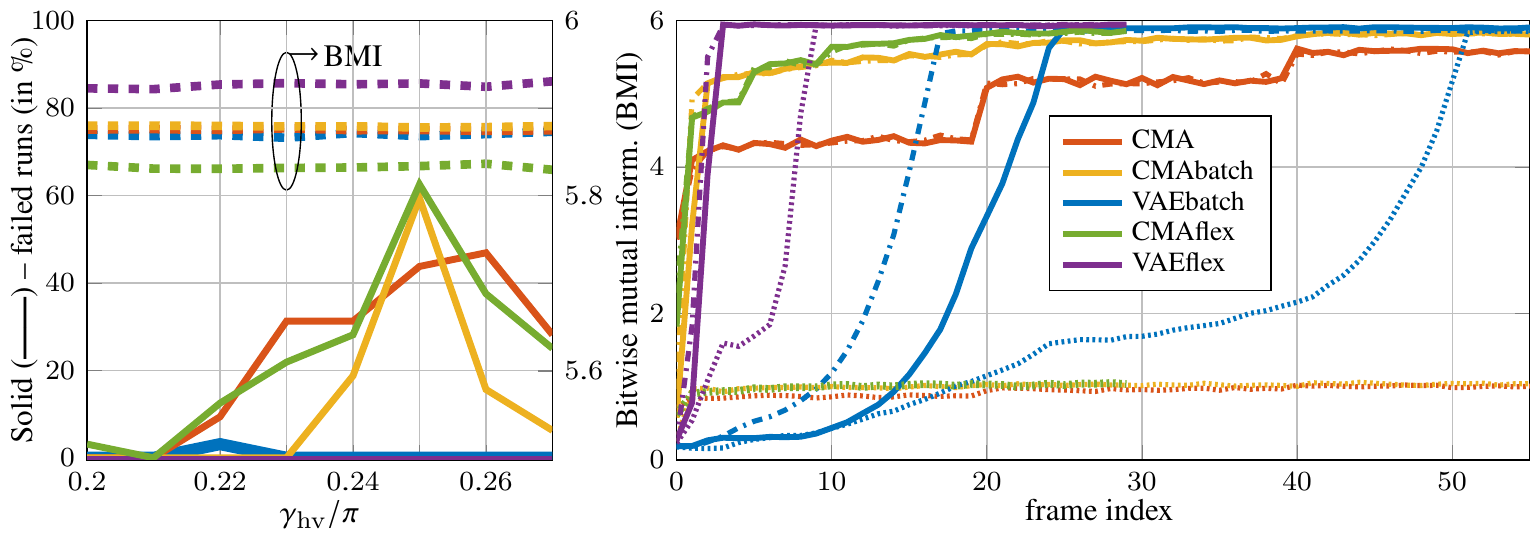}
		\vspace*{-3ex}
		\caption{Results for different initial HV-shift $\gammahv$ (left), and the convergence behavior at $\gammahv=\num{0.25}\pi$ (right), where the solid (-----) curve corresponds to the convergence of a typical run, the dotted ($\mathbf{\cdot\!\cdot\!\cdot\!\cdot\!\cdot\!}$) to the convergence of a bad run, and the dash-dotted ($\mathbf{\cdot-\cdot}$) to the convergence of a good run. }
		\label{Results:pol}
		\vspace*{-4ex}
\end{figure} 
\vspace*{-1ex}
\section{Numerical Results}\vspace*{-1.5ex} 
All numerical results are simulated with $\Meq=15$ and with a symbol rate of $R_{\text{S}}=\SI{100}{\giga Bd}$, an oversampling factor of $N_\text{os}=2$~\ac{sps}, and an SNR of $\SI{24}{dB}$ for uniform and $\SI{22}{dB}$ for \ac{PCS}-$64$-QAM with $\mathcal{H}=5.73$. 
 
First, we analyze the convergence behavior at a challenging working point with $\gammahv=\num{0.2}\pi$, $\taupmd = \frac{T_\text{S}}2 = \frac1{2 R_\text{S}}$, and for various lengths $\Lcd$ of uncompensated fiber as depicted in Fig.~\ref{Results:Lcd_full}. To take the potentially long channel impulse response into account, we use $\Mest=25$~taps. 
While the \ac{VAE} equalizers achieve the highest \ac{BMI}, the CMAbatch is the most stable algorithm for uniform QAM, where it converges reliably and within 5 frames as depicted in the right plot. 
The CMAflex/VAEflex converge as fast as the CMAbatch, but we observe a higher percentage of failed runs (left plots) for higher \ac{CD}. However, the VAEflex achieves a significantly higher BMI than the CMAbatch and CMAflex. The computationally less expensive VAEbatch algorithm converges similarly reliably and equalizes well, as indicated by the constellation diagram in Fig.~\ref{Results:Lcd_full} as well as the high BMI. 
The bottom row depicts the results for moderately shaped \ac{PCS}-64-QAM with an entropy of $\mathcal{H}=5.73$, which is commonly-known as challenging task for the popular \ac{CMA} based blind equalizers. Only the VAE based equalizers, which consider the \textit{a~priori} density, converge reliably until the \ac{CD} becomes too strong, while the CMAflex fails completely. The VAEflex is still converging within 10 frames but the VAEbatch converges more reliably. 

At $\gammahv=\frac{\pi}4$, the \ac{CMA}'s loss function has a local minimum at an invalid constellation, so the left plot of Fig.~\ref{Results:pol} depicts a sweep of $\gammahv$ around that point. We set a rather strong \ac{PMD} with $\taupmd=T_\text{S}$, a moderate $\Lcd=\SI{1}{\kilo\meter}$, and use $\Mest=15$. While the BMI after convergence stays similar (dashed), the percentage of failed runs (solid) increases significantly at $\gammahv\approx\frac{\pi}4$  for all \ac{CMA} based equalizers, which eventually fail in over 50\% of the runs. However, the VAE based equalizers do not struggle at all for this working point. 
The right plot shows the convergence behavior at the critical point of $\gammahv=\frac{\pi}4$. We chose three runs per algorithm representing a good (dash-dotted) and a bad (dotted) performance as well as typical curve (solid) of the converged runs. While the \ac{CMA} based equalizers quickly reach a moderate BMI and converge slowly but gradually, the VAE based algorithms start at low BMI but converge rapidly. Hence, they reach their maximum BMI faster as all CMA based equalizers. 
The results show that all algorithms are significantly affected by the learning rate scheduler, which might be subject of further optimization.

\vspace*{-1ex}
\section{Conclusion}\vspace*{-1.5ex}
We reveal critical working points in which the widely-used \ac{CMA} and its variants fail to converge. The novel \ac{VAE} based equalizers converge reliably at these critical points with both standard and \ac{PCS} formats. 
The novel VAEflex is a fast-converging option and, for the CMA, we found that batch-wise updating is advantageous. 

\vspace*{2ex}

\footnotesize 
\noindent{\bf{Acknowledgements}}:
This work was carried out in the framework of
the CELTIC-NEXT project AI-NET-ANTILLAS (C2019/3-3) and was funded
by the German Federal Ministry of Education and Research (BMBF) under
grant agreement 16KIS1316.

\end{document}